\documentclass[12pt]{article}

\pdfoutput=1

\usepackage{cite}

\usepackage{amssymb,latexsym}
\usepackage{amsmath}

\usepackage{epstopdf}

\usepackage{graphics,epsfig}

\usepackage{color}

\usepackage{mathrsfs}

\DeclareMathAlphabet{\mathpzc}{OT1}{pzc}{m}{it}

\usepackage{tikz}
\usetikzlibrary{arrows,shapes}
\usetikzlibrary{trees}
\usetikzlibrary{matrix,arrows} 				
\usetikzlibrary{positioning}				
\usetikzlibrary{calc,through}				
\usetikzlibrary{decorations.pathreplacing}  
\usepackage{pgffor}							

\usetikzlibrary{decorations.pathmorphing}	
\usetikzlibrary{decorations.markings}
\tikzset{
    vector/.style={decorate, decoration={snake}, draw},
	provector/.style={decorate, decoration={snake,amplitude=2.5pt}, draw},
	antivector/.style={decorate, decoration={snake,amplitude=-2.5pt}, draw},
    fermion/.style={draw=black, postaction={decorate},
        decoration={markings,mark=at position .55 with {\arrow[draw=black]{>}}}},
    fermionbar/.style={draw=black, postaction={decorate},
        decoration={markings,mark=at position .55 with {\arrow[draw=black]{<}}}},
    fermionnoarrow/.style={draw=black},
    gluon/.style={decorate, draw=black,
        decoration={coil,amplitude=4pt, segment length=5pt}},
    scalar/.style={dashed,draw=black, postaction={decorate},
        decoration={markings,mark=at position .55 with {\arrow[draw=black]{>}}}},
    scalarbar/.style={dashed,draw=black, postaction={decorate},
        decoration={markings,mark=at position .55 with {\arrow[draw=black]{<}}}},
    scalarnoarrow/.style={dashed,draw=black},
    electron/.style={draw=black, postaction={decorate},
        decoration={markings,mark=at position .55 with {\arrow[draw=black]{>}}}},
	bigvector/.style={decorate, decoration={snake,amplitude=4pt}, draw},
}

\tikzstyle{block} = [draw, rectangle, 
    minimum height=3em, minimum width=6em]

\let\a=\alpha \let\b=\beta \let\g=\gamma \let\d=\delta \let\e=\epsilon
\let\z=\zeta  \let\th=\theta  \let\k=\kappa
\let\l=\lambda \let\m=\mu \let\n=\nu \let\x=\xi \let\p=\pi 
\let\s=\sigma \let\t=\tau  \let\f=\phi  
 
\let\w=\omega      \let\G=\Gamma  \let\Th=\Theta \let\L=\Lambda
\let\X=\Xi  \let\S=\Sigma  \let\Y=\Psi
 
\let\la=\label  
  
\def\nn{\nonumber} \def\bd{\begin{document}} \def\ed{\end{document}}
\def\ds{\documentstyle} \let\fr=\frac \let\bl=\bigl \let\br=\bigr
\let\Br=\Bigr \let\Bl=\Bigl
\let\bm=\bibitem
\let\na=\nabla
\def\tU{{\widetilde U}}
\let\pa=\partial \let\ov=\overline
\def\ie{{\it i.e.\ }}
\newcommand{\be}{\begin{equation}}
\newcommand{\ee}{\end{equation}}
\def\ba{\begin{array}}
\def\ea{\end{array}}
\def\ft#1#2{{\textstyle{{\scriptstyle #1}\over {\scriptstyle #2}}}}
\def\fft#1#2{{#1 \over #2}}
\def\F#1#2{{ F_{#1}^{(#2)} }}
\def\cF#1#2{{ {\cal F}_{#1}^{(#2)} }}

\def\R{{\bf R}}
\def\sst#1{{\scriptscriptstyle #1}}
\def\oneone{\rlap 1\mkern4mu{\rm l}}
\def\e7{E_{7(+7)}}
\def\td{\tilde}
\def\wtd{\widetilde}
\def\im{{\rm i}}
\def\bog{Bogomol'nyi\ }
\newcommand{\ho}[1]{$\, ^{#1}$}
\newcommand{\hoch}[1]{$\, ^{#1}$}
\newcommand{\bea}{\begin{eqnarray}}
\newcommand{\eea}{\end{eqnarray}}
\newcommand{\ra}{\rightarrow}
\newcommand{\lra}{\longrightarrow}
\newcommand{\Lra}{\Leftrightarrow}
\newcommand{\ap}{\alpha^\prime}
\newcommand{\bp}{\tilde \beta^\prime}
\newcommand{\cB}{{\cal B}}
\newcommand{\cO}{{\cal O}}
\newcommand{\vecx}{\vec{x}}
\newcommand{\vecy}{\vec{y}}
\newcommand{\vecp}{\vec{p}}
\newcommand{\vecq}{\vec{q}}
\newcommand{\tr}{{\rm tr} }
\newcommand{\Tr}{{\rm Tr} }
\newcommand{\NP}{Nucl. Phys. }

\newcommand{\cL}{{\cal L}}
\newcommand{\cA}{{\cal A}}
\newcommand{\cT}{{\cal T}}
\newcommand{\cR}{{\cal R}}
\newcommand{\cD}{{\cal D}}
\newcommand{\cH}{{\cal H}}

\def\Cb{\bar{C}}

\def\sst#1{{\scriptscriptstyle #1}}
\def\0{{\sst{(0)}}}
\def\1{{\sst{(1)}}}
\def\2{{\sst{(2)}}}
\def\3{{\sst{(3)}}}
\def\4{{\sst{(4)}}}
\def\5{{\sst{(5)}}}
\def\6{{\sst{(6)}}}
\def\7{{\sst{(7)}}}
\def\8{{\sst{(8)}}}
\def\9{{\sst{(9)}}}
\def\p{{\sst{(p)}}}
\def\q{{\sst{(q)}}}
\def\ve{\varepsilon}
\def\vf{\varphi}
\def\F{\Phi}
\def\wg{\wedge}

\def\thb{\bar{\theta}}
\def\Thb{\bar{\Theta}}
\def\barp{\bar{p}}
\def\barq{\bar{q}}
\def\barc{\bar{c}}
\def\bard{\bar{d}}
\def\e{\epsilon}

\def \bi{\bibitem}
\def \la {\label}

\def \l {\lambda}
\def\foot{\footnote}
\def \tl  {{\tilde \l}}
\def \sql {{\sqrt \l}}
\def \adss {$AdS_5 \times S^5$\ }
\newcommand{\rf}[1]{(\ref{#1})}
\def \ov {\over}

\def\th{\theta}
\def\Th{\Theta}
\def\vth{\vartheta}
\def\btheta{{\bar\theta}}
\def\ttheta{{{\tilde\theta}}}
\def\bttheta{{{\bar\ttheta}}}
\def\vth{\vartheta}

\def\ra{\rightarrow}
\def\N{\nabla}
\def\F{{\cal F}}
\def\uM{\underline{M}}
\def\uA{\underline{A}}
\def\uN{\underline{N}}
\def\uP{\underline{P}}
\def\ua{\underline{a}}
\def\ub{\underline{b}}
\def\uc{\underline{c}}
\def\ud{\underline{d}}
\def\ue{\underline{e}}
\def\uf{\underline{f}}
\def\ui{\underline{i}}
\def\uj{\underline{j}}
\def\uk{\underline{k}}
\def\ul{\underline{l}}
\def\ual{\underline{\alpha}}
\def\ube{\underline{\beta}}
\def\um{\underline{m}}
\def\un{\underline{n}}
\def\up{\underline{p}}
\def\uq{\underline{q}}
\def\ur{\underline{r}}
\def\us{\underline{s}}
\def\umu{\underline{\mu}}
\def\unu{\underline{\nu}}
\def\ula{\underline{\l}}
\def\uka{\underline{\k}}
\def\usi{\underline{\s}}
\def\urh{\underline{\r}}
\def\cc{\circ}
\def\eqv{\equiv}

\def\ni{\noindent}

\def\Ep{E^{{}^{(+)}}}
\def\Em{E^{{}^{(-)}}}

\def\Mp{M^{{}^{(+)}}}
\def\Mm{M^{{}^{(-)}}}

\def \ha{{1\ov 2}}

\def\r{\rho}

\def\Y{{\rm Y}}
\def\X{{\rm X}}
\def\tY{\tilde{\rm Y}}
\def\tX{\tilde{\rm X}}
\def\dY{\dot{\rm Y}}
\def\dX{\dot{\rm X}}

\def \J {\mathcal{J}}
\def \del {\partial}

\def\dF{\dot{F}}
\def\dG{\dot{G}}
\def\df{\dot{f}}
\def \E {{\cal E}}
\def \S {{\cal S}}
\def \J {{\cal J}}

\def\ms{\mathcal{S}}
\def\mj{\mathcal{J}}
\def\soj{\fr{\ms}{\mj}}
\def \R {{\bf R}}
\def \om {\omega}
\def \bE {\bar E}
\def \x {{\cal X}}

\def \bi{\bibitem}
\def \la {\label}

\def \l {\lambda}
\def\foot{\footnote}
\def \tl  {{\tilde \l}}
\def \sql {{\sqrt \l}}
\def \adss {$AdS_5 \times S^5$\ }
\def \ov {\over}

\def \varpi {{\rm w}}

\def\thb{\bar{\theta}}
\def\Thb{\bar{\Theta}}
\def\mb{\bar{\m}}
\def\ab{\bar{\a}}
\def\zb{\bar{z}}
\def\psib{\bar{\psi}}
\def\barp{\bar{p}}
\def\barq{\bar{q}}
\def\barc{\bar{c}}
\def\bard{\bar{d}}
\def\e{\epsilon}
\def\wb{\bar{w}}
\def\lb{\bar{\l}}
\def\Jb{\bar{J}}
\def\Nb{\bar{N}}
\def\Zb{\bar{Z}}
\def\pab{\bar{\pa}}

\def\ab{{\bar{a}}}
\def\Ab{\bar{A}}
\def\pb{\bar{p}}
\def\qb{\bar{q}}
\def\cb{\bar{c}}
\def\db{\bar{d}}
\def\vb{\bar{v}}
\def\ub{\bar{u}}
\def\zb{\bar{z}}
\def\Zb{\bar{Z}}

\def\At{\tilde{A}}
\def\Bt{\tilde{B}}
\def\Ct{\tilde{C}}
\def\Dt{\tilde{D}}
\def\Et{\tilde{E}}
\def\Ft{\tilde{F}}
\def\Gt{\tilde{G}}
\def\Ht{\tilde{H}}
\def\Kt{\tilde{K}}
\def\Mt{\tilde{M}}
\def\Nt{\tilde{N}}
\def\Rt{\tilde{R}}
\def\at{\tilde{a}}
\def\bt{\tilde{b}}
\def\ct{\tilde{c}}
\def\dt{\tilde{d}}
\def\et{\tilde{e}}
\def\ft{\tilde{f}}
\def\htil{\tilde{h}}
\def\gt{\tilde{g}}
\def\nt{\tilde{n}}
\def\mut{\tilde{\mu}}
\def\nut{\tilde{\nu}}
\def\pht{\tilde{\f}}
\def\Pht{\tilde{\Phi}}
\def\vft{\tilde{\vf}}
\def \zet{\tilde{\z}}

\def\rht{\tilde{\rho}}

\def\asth{\hat{*}}
\def\phh{\hat{\phi}}

\def\bA{{\bf A}}

\def\ola{\overleftarrow}
\def\ora{\overrightarrow}
\def\alt{\tilde{\a}}

\def\eh{\hat{e}}
\def\eph{\hat{\e}}
\def\ph{\hat{p}}
\def\alh{\hat{\a}}
\def\beh{\hat{\b}}
\def\gah{\hat{\g}}
\def\Fh{\hat{F}}
\def\muh{\hat{\m}}
\def\nuh{\hat{\n}}
\def\thh{\hat{\th}}
\def\rhh{\hat{\r}}
\def\dh{\hat{d}}
\def\ih{\hat{i}}
\def\jh{\hat{j}}
\def\hh{\hat{h}}
\def\nh{\hat{n}}
\def\gh{\hat{g}}
\def\kh{\hat{k}}
\def\deh{\hat{\d}}
\def\wh{\hat{w}}
\def\lah{\hat{\l}}
\def\rh{\hat{r}}
\def\Ah{\hat{A}}
\def\Kh{\hat{K}}
\def\Nh{\hat{N}}
\def\Rh{\hat{R}}
\def\Ch{\hat{C}}
\def\Omh{\hat{\Omega}}

\def\xh{\hat{x}}

\def\ps{\rlap{\, /}\;\,p }
\def\ks{\rlap{\, /}\;\,k }

\def\gym{g_{YM}}

\def\adot{\dot{a}}
\def\bdot{\dot{b}}
\def\bpa{\bar{\pa}}

\def\pr{\prime}
\def\ssk{\medskip}
\def\clb{\color{blue}}
\def\clr{\color{red}}
\def\clg{\color{green}}

\def\bfA{{\bf A}}
\def\bfB{{\bf B}}
\def\bfK{{\bf K}}
\def\bfU{{\bf U}}
\def\bfX{{\bf X}}
\def\bfY{{\bf Y}}
\def\bfZ{{\bf Z}}
\def\bfg{{\bf g}}
\def\bfn{{\bf n}}

\def \vk{\vec{k}}
\def \vx{\vec{x}}

\def\kbf{\mathbf{k}}
\def\pbf{\mathbf{p}}
\def\ybf{\mathbf{y}}
\def\xbf{\mathbf{x}}
\def\Lbf{\mathbf{L}}
\def\rbf{\mathbf{r}}

\begin{document}

\overfullrule=0pt
\parskip=2pt
\parindent=12pt
\headheight=0in \headsep=0in \topmargin=0in
\oddsidemargin=0in

\vspace{ -3cm}
\thispagestyle{empty}

 \vspace{0.1cm}

\setcounter{equation}{0}
\setcounter{footnote}{0}
\setcounter{section}{0}

\begin{center}

{\Large\bf  Quantum-corrected geometry of horizon vicinity}

\vskip 0.8cm

 \vspace{.5cm}

\vspace{0.5cm}
I. Y. Park
\\

\vspace{0.3cm}

\vspace{0.3cm}
{\it Department of Applied Mathematics,
Philander Smith College 
                               \\
Little Rock, AR 72202, USA \\
inyongpark05@gmail.com
}

\end{center}

 \vspace{0.1cm}

\begin{abstract}

We study the deformation of the horizon-vicinity geometry caused by quantum gravitational effects. Departure from the semi-classical picture is noted, and the fact that the matter part of the action comes at a higher order in Newton's constant than does the Einstein-Hilbert term is crucial for the departure. The analysis leads to a Firewall-type energy measured by an infalling observer for which quantum generation of the cosmological constant is critical. The analysis seems to suggest that the Firewall should be a part of such deformation and that the information be stored both in the horizon-vicinity and asymptotic boundary region. We also examine the behavior near the cosmological horizon.

\end{abstract}
\newpage

\section{Introduction}

With the recent experimental confirmation of the gravitational wave \cite{Abbott:2016blz}, an exciting time of gravitational astrophysics lies ahead.
The quantum gravitational effects \cite{DeWittBook} are expected to play a major role in strong gravitational astrophysical environments such as near a black hole.  
The potential importance of the quantum gravitational effects in black hole information \cite{Hawking:1976ra} (see, e.g., \cite{Page:1993up,Mathur:2009hf,tHooft:1996rdg,Hooft:2016vug,Dvali:2012rt,Polchinski:2016hrw,Marolf:2017jkr,Unruh:2017uaw} for reviews and various viewpoints) has been emphasized in \cite{Park:2013rm}. 
One of the main goals of the present work is to further substantiate the proposal in \cite{Park:2013rm,Park:2014mba}, put forth in support of Firewall \cite{Almheiri:2012rt}\cite{Braunstein:2009my,Braunstein:2014nwa} (see \cite{Itzhaki:1996jt,Mazur:2001fv,Chapline:2000en,Czech:2012be} also for other related ideas), that an infalling object should experience a non-smooth entry through the horizon due to the loop effects: an infalling observer or an incoming wave-packet will experience the effects of the multi-particle emission (called the ``jets"\footnote{For a more realistic black hole, the inner region of the accretion disk should be the main region for manifestation of multi-jet activities. Other interesting possibilities include the recent models of absence of the horizon \cite{Kawai:2013mda} \cite{Ho:2017joh} \cite{Moskalets:2016uno}.} in \cite{Park:2013rm}) initiated by the infalling entity.

In recent works \cite{Park:2016fxc,Park:2016vam} (see, e.g., \cite{Freidel:2016bxd,Donnelly:2016auv} for related issues), it has been shown that the quantum effects influence the boundary conditions of a gravitational system. (Perhaps this might be related to the observation made in \cite{Czech:2012be}.) Based on \cite{Park:2013rm}, the bulk geometry is also expected to be affected: the quantum effects should modify the classical black hole geometry (an effect that may perhaps be in line with the ``quantum atmosphere" \cite{Giddings:2017mym}\cite{Dey:2017yez}). As a matter of fact, the renormalization analysis in \cite{Park:2016zgt} has revealed the deformation of the bulk metric, as implied by the field redefinition of the metric: the quantum corrected solution signifies the deformation of the geometry by loop effects.

Although it should be possible to directly study the multi-particle emissions of an incoming particle, a very technically demanding task, there should be other simpler ways to probe the quantum effects to the geometry and phenomena experienced by an infalling observer. One such route would be to analyze the energy density measured by an infalling observer \cite{Lowe:2013zxa}\cite{Park:2014mba}\cite{Mann:2014yxa}. 
We employ the setup of \cite{Park:2016zgt} and show that an infalling observer experiences an unsmooth entry through the horizon and measures a trans-Planckian energy at the horizon, although the stress-energy tensor is finite. The Firewall-like behavior has its origin in the quantum-gravity-induced cosmological constant as the foremost factor.

The computation of stress energy has a long history (see, e.g., \cite{Unruh:1976db,Fulling:1977zs,Christensen:1977jc,Candelas:1980zt,Birrell}). Due to various obstacles, the analyses took rather indirect approaches in the past. For example, most of the analyses employed a system with conformal symmetry so that the form of the stress tensor could be constrained. The presence of a large amount of gauge symmetry and its fixing (see, e.g., the discussions in \cite{Kuchar:1970mu,Gibbons:1978ac,Barvinsky:1985an,Schleich:1987fm,Mazur:1989by}) has also been slowing progress in more direct analysis. However, there has been recent progress in both the gauge-fixing procedure \cite{Park:2014tia} and background field Feynman diagrammatic techniques \cite{Park:2015ota}. In this work we consider a gravity-scalar system without conformal symmetry, and those results are utilized to analyze and deduce the generation of the cosmological constant through loop effects in various curved backgrounds. The presence of the cosmological constant will play an important role in the trans-Planckian energy measured by an infalling observer in the time-dependent black hole background obtained in \cite{Chadburn:2013mta} (earlier related works can be found in \cite{Jacobson:1999vr} and \cite{Frolov:2002va}).

The quantum gravitational effects should be present in any gravitational system and therefore it ought to be possible to demonstrate them generically. However, choosing a simple but still realistic system will greatly reduce the amount of loop calculation required.
For reasons to be explained in the main body, a time-dependent configuration provides a setup that requires less loop computation. By invoking the loop effects in the time-dependent background of \cite{Chadburn:2013mta}, we show below that its {\em quantum-corrected} geometry supports a Firewall-type energy and related unsmooth horizon ideas: although the stress-energy tensor itself is regular at the horizon, the energy measured by an infalling observer becomes trans-Planckian.

The analysis in the main body consists of several components. As we will see in section 3, the presence of the cosmological constant is critical for the leading-order metric back-reacted geometry obtained in \cite{Chadburn:2013mta}, the background that we employ in order to be specific. In the first part of the analysis that we take up in section 2, we investigate the loop-induced generation of the cosmological constant. Basically, what we establish is the relevance of the quantum gravitational effects that occur generically, i.e., regardless of the backgrounds considered and naturalness of using the renormalized value of the cosmological constant when it comes to solving the quantum corrected field equations. 
With that established we turn in section 3 to analyzing the significance of the cosmological constant in constructing the time-dependent solution \cite{Chadburn:2013mta}, which in turn is crucial in establishing the trans-Planckian energy.    
The loop analysis via use of the actual propagator associated with the background of \cite{Chadburn:2013mta} would be extremely involved.
For this reason we employ a Schwarzschild geometry for the analysis in section 2 where the genericity of the loop-generation of cosmological constant is illustrated.
 It is this genericity of the quantum effects that we rely on to draw certain conclusions about the time-dependent background, \cite{Chadburn:2013mta}, based on consideration of a relatively simple background such as a flat or Schwarzschild.

In general, one of the central issues in a curved space Feynman diagramatic analysis is the regularization method: the loop correction terms are ultraviolet-divergent and requires regularization. Although the regularization would be complicated for a generic Feynman diagram, we are interested in evaluating {\em vacuum} diagrams, and in that case it is possible to avoid the complexity. Also, when perturbatively computing the loop-induced coefficients of the various terms in the effective action, the coefficients are expected, in general, to depend on the parameter of the Schwarzschild geometry (or the geometry under consideration), the mass of the black hole. Interestingly, however, it turns out that the coefficient (including the finite part) of the cosmological constant does not depend on the mass of the black hole; instead, it is purely numerical.

What brings the departure from the semi-classical picture is the relative weights between the classical and quantum contributions: the conditions for smallness of quantum gravitational effects in the literature (see, e.g., \cite{Mathur:2009hf} for a review) does not render small, compared with the classical matter contributions, the effect associated with the generation of the cosmological constant by quantum effects. This is because one-loop gravity vacuum diagram comes at the same order as the tree-level matter fields: the contribution of one-loop vacuum diagram to the stress energy tensor is comparable to the (semi-) classical matter contributions. As far as we can tell, this should be where some of the assumptions of the semi-classical framework are not fully justified.

\vspace{.3in}

The rest of the paper is organized as follows.

\vspace{.1in}

In general the Green's function for a curved background is a very complicated object.
Because the full analysis would be highly technical, we start in section 2 by considering the easier case of Schwarzschild geometry in order to outline the strategy.\footnote{The goal of section 2 is to set the stage for the rationale (to be established in section 3) that the divergent energy measured by an infalling observer should be attributed to the quantum effect. Although it is a crucial component of our analysis, the details of the loop computation are not needed for the analysis in section 3.} Although the loop analysis of the one-loop vacuum diagram is much simpler than the case of the time- and position- dependent background of \cite{Chadburn:2013mta}, it is still nontrivial because the regularization and the Green's function for the Schwarzschild geometry are complex. Fortunately, however, it is possible, for the vacuum diagrams at least, to bypass these complications by making a clever choice of the basis when taking the trace.
By taking the Higgs-type system to be specific, we demonstrate generation of the cosmological constant from quantum effects and outline the renormalization procedure of the cosmological constant. (See, e.g., \cite{Sola:2013gha} for a review of renormalization of the cosmological constant in the context of the cosmological constant problem.) We also point out that the metric field redefinition required as part of the renormalization procedure \cite{Park:2016zgt} implies that the geometry gets deformed by the quantum effects.  
In section 3, we review the time- and position- dependent solution of \cite{Chadburn:2013mta} constructed on general grounds. The background is viewed as a solution of the renormalized action with the quantum corrections and we compute the energy measured by an infalling observer.
For the purpose of demonstrating the trans-Planckian energy measured by an infalling observer, we focus on the matter (i.e., scalar) kinetic term for two reasons. First of all, we are not sure whether there is consensus on whether the Riemann tensor-containing quantum-correction terms should be included in the matter sector or gravity sector.\footnote{ The reasonable thing to do is to include in the stress-energy tensor only those higher curvature terms arising from the {\em matter sector}, i.e., with the matter fields running in the loop.} Secondly, obviously the kinetic term is simple, and if the sought-for quantum effects can be demonstrated thereby (which we show is the case), that will make the analysis less involved than examination of the Riemann tensor-containing terms (which, in any case, are sub-leading in the derivative expansion). The analysis leads to a trans-Planckian energy measured by an infalling observer and is in line with the Firewall and related ideas. 
Since the loop effects will generate and renormalize the cosmological constant term, the trans-Planckian energy measured by an observer must be taken as a quantum effect. In the conclusion, we summarize and discuss several implications of the results. It is noted that the information is stored split over the horizon vicinity and asymptotic region, worth more detailed study.

\section{Graviton quantum-field-theoretic effects}

The quantity to be computed, ultimately, is the energy density measured by a free-falling observer:
\bea
T_{\m\n}^{quan}U_K^\m U_K^\n  \la{2dse}
\eea
where $U_K^\r$ denotes the four-velocity of an infalling observer in the Kruskal coordinates. $T_{\m\n}^{quan} \equiv <K|\;T_{\m\n}^K \;|K>$ denotes the quantum-corrected stress tensor where $|K>$ is the Kruskal vacuum (i.e., Hartle-Hawking vacuum). One approach, which we follow, to compute 
$T_{\m\n}^{quan} $ is to first compute the one-loop 1PI action followed by taking a metric variation of the matter part. To our knowledge, this approach of computing the stress tensor has not been explicitly attempted in the literature even for a Schwarzschild background, let alone a more complicated background. The one-loop renormalizability was explicitly established in \cite{Park:2016zgt} for a gravity-scalar system.\footnote{What has made the difference compared with the analyses in the past is the inclusion of the cosmological constant; more in the conclusion.} Although the (all-loop) renormalizability was established for a certain class of backgrounds, we do not, at one-loop, expect the time- and position- dependence to cast any additional difficulty to the matters of principle. This is because at the one-loop level, the appearance of Riemann tensors - which was the source for the non-renormalizability in higher loops - can be controlled without employing the reduction device of \cite{Park:2014tia}.
The metric variation of the matter part of the resulting effective action will yield the one-loop stress-energy tensor. 
Although the procedure is conceptually straightforward, the analysis requires highly technical calculations due to the complexities in regularization and the Green's function in the time-dependent background. For this reason we analyze the cases of the flat and Schwarzschild backgrounds to demonstrate the genericity of the loop-generated cosmological constant.

After reviewing generation of the cosmological constant in a flat background in section 2.2, we analyze in section 2.3 how the loop effects generate the cosmological constant term in the Schwarzschild background. From these analyses, one can see the generic character of the  loop-generation of the cosmological constant.  
We make an assertion based on this genericity on the occurrence of the cosmological constant in a generic background including the time-dependent black hole background of \cite{Chadburn:2013mta}.

The gravity-scalar system that we consider is
\bea
S=\fr1{\k_r^2}\int d^4 x \sqrt{-g_r}\; R_r  - \int d^4 x\sqrt{-g_r}\; \Big(\fr12g_r^{\m\n}\pa_\m\f_r \pa_\n \f_r +{\cal V}_r\Big)
\la{grv-sclrq3}
\eea
\bea
{\cal V}_r &=& \fr{\l_r}{4}\Big(\f_r^2+\fr{1}{\l_r} \m_r^2\Big)^2   \label{potqm}
\eea
where the subscript $r$ indicates the renormalized quantities; it will be suppressed below. 
Taking the Higgs-type system not only makes the analysis more specific, but also leads to more realistic phenomenological implications.

\subsection{motivation}

Our motivation for considering the quantum effects comes from the picture presented in \cite{Park:2013rm}.
Suppose there is a wavepacket infalling on an otherwise classically stationary black hole. Just as one would analyze the multi-particle emission amplitudes in a flat background, it should be possible, although technically much more laborious, to calculate analogous amplitudes in the black hole background under consideration. Due to the presence of various vertices in the Lagrangian, the incoming packet will produce multi-particles with various angles with respect to the direction toward the black hole. Conventionally such quantum gravitational effects are asserted to be small, and this must be the rationale for the ``information-free" horizon and ``no-hair" theorem. However, as we will show in section 3, there are circumstances in which this assumption may not be justified. It is not only that the assumption may not be justifiable but also that the circumstance in which it is not justifiable is actually a very realistic one. Some of those subsequent particles will be directed away from the black hole, reaching infinity, whereas others will more actively contribute to further production of multi-particles as schematically depicted in Fig. 1. Such chain reactions will lead to an environment of the horizon-vicinity that is quite different from the semi-classical picture.

\begin{figure}
\centerline{
\begin{minipage}[b]{12cm}
             \epsfxsize=14cm
              \epsfbox{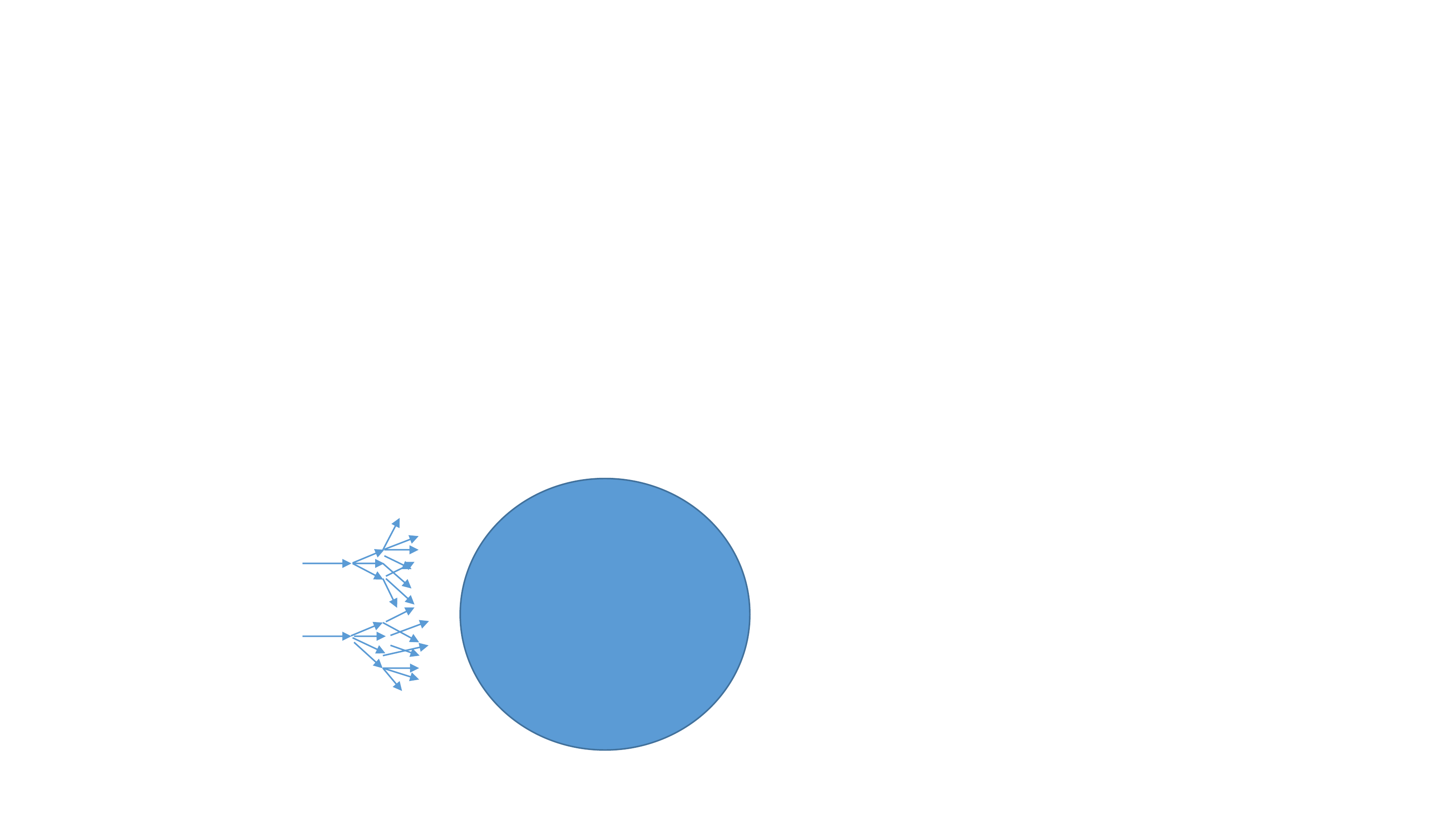}
      \end{minipage}
      }
\caption{jet production in the horizon vicinity}
\label{fig1}
\end{figure}

Evidently the analysis of the multi-particle emission process will be technically challenging though not impossible. In the present work we take a closer look at what should be a much easier task of computing the energy measured by an infalling entity. Since the horizon-vicinity should potentially be a volatile region under small perturbation, it seems natural to anticipate that the infalling entity will undergo a violent entry as it approaches the horizon.

\subsection{loop-induced cosmological constant}

In spite of the lack of a closed-form expression for a curved spacetime propagator in general, certain things can be deduced from consideration of simpler backgrounds. As we will see below, this includes one-loop vacuum diagrams. One of the messages that we want to convey with the analyses of the present and next subsections is that the appearance of the cosmological constant through the loop effects is quite generic. The common lore, confirmed in our previous works (see, e.g., \cite{Park:2016zgt}), that the divergence can be computed by considering a flat spacetime is confirmed again. 
In this subsection, we analyze the loop-induced generation of the cosmological constant by starting with a flat background. As we will see in section 3, the presence of the cosmological constant will be crucial for the leading time-dependence of the solution found in \cite{Chadburn:2013mta}, which in turn will be important later for the trans-Planckian energy measured by an infalling observer.

Needless to say, maintaining the 4D covariance offers a great advantage in establishing renormalizability. (The analysis of the quantum gravitational effects is inevitably tied with the renormalizability issue which we therefore frequently visit.) 
In gravitational theories, more care is needed to find the counterterms in a covariant manner: given that the action is expanded around the background under consideration, maintaining the covariance is more subtle compared with the non-gravitational gauge theories. As analyzed in detail in \cite{Park:2015ota}, the covariance can be achieved by the following double shift:
\bea
g_{\m\n}\equiv  h_{\m\n}+g_{{}_B\m\n}+g_{0\m\n}
\eea
and basically treating $\gt_{\m\n}\equiv g_{{}_B\m\n}+g_{0\m\n}$ as a single piece, which then allows one to have a perturbative series set up with the propagator associated with the background $g_{0\m\n}$. As a matter of fact, the basically same shift had been considered in \cite{Kallosh:1978wt,Capper:1984qq,Antoniadis:1995fc}. It is a double shift in the sense that the metric $g_{\m\n}$ is initially shifted around the background $g_{0\m\n}$:
\bea
g_{\m\n}\equiv  h_{\m\n}+g_{0\m\n}
\eea
followed by 
\bea
h_{\m\n} \ra h_{\m\n}+g_{{}_B\m\n}
\eea
The double shift allows one to perturbatively compute the effective action around the background $g_{0\m\n}$ in the background field method.

Let us narrow down to the computation of the generation of the cosmological constant.
As shown in \cite{Park:2016zgt}, the analysis required to show the generation and renormalization of the cosmological constant is a curved space generalization of the standard calculation (see, e.g., \cite{Weinberg2}). Let us briefly illustrate the computation by taking the Einstein-Hilbert sector; more details can be found in \cite{Park:2016zgt}.
After multiplying by an overall factor two, the kinetic term of the Einstein-Hilbert action is given by
\bea
\cL&=& -\fr12 {\pa}_\g h^{\a\b}{\pa}^\g h_{\a\b}+\fr14 {\pa}_\g h^{\a}_\a {\pa}^\g h^{\b}_\b  
+ m^2 \Big(-\fr12 h_{\m\n}h^{\m\n}+\fr14 h^2 \Big) \la{eawm}
\eea
where
	\bea
	\ m^2 \equiv  \fr14 \fr{\k^2 \m^4}{\l}       \la{amass}
	\la{gmass}
	\eea
The parameter $\m$ appeared in \rf{potqm}.	
When expanded, the quadratic order of the constant piece of the scalar potential yields
	\bea
	\fr14 m^2\int  (h^2-2 h_{\m\n}h^{\m\n}) \la{gmassterm}
	\eea
The term \rf{gmassterm} can be combined with the original massless propagator of the graviton. The propagator then is given by
\bea 
<\f_{\m\n}(x_1)\f_{\r\s}(x_2)>&=& P_{\m\n\r\s} \; (2\k^2)\int \fr{d^4k}{(2\pi)^4}\fr{e^{ik\cdot (x_1-x_2)}}{i (k^2+m^2)} 
\eea
where $P_{\m\n\r\s}$ for the traceless propagator takes\footnote{The propagator-dependence of the effective action was analyzed, e.g., in \cite{Odintsov:1990qq}.}:
\bea
P_{\m\n\r\s}\equiv \fr12(\eta_{\m\r}\eta_{\n\s}+\eta_{\m\s}\eta_{\n\r}- \fr12\eta_{\m\n}\eta_{\r\s})
\eea
where $\eta_{\m\n}=diag(-1,1,1,1)$.
Note that $P_{\m\n\r\s}$ is traceless in $(\m\n)$ and similarly for $(\r\s)$. The traceless gauge-fixing of the fluctuation metric was noted in \cite{Ortin} some time ago. See also the work of \cite{Kuchar:1970mu} for an earlier related discussion.
The contribution leading to the cosmological constant renormalization comes from
\bea 
\int \prod_x dh_{\k_1\k_2}\;e^{\fr{i}2 \int \sqrt{\gt} \;h^{\a\b}({\pa}^2-m^2) h_{\a\b}  }
\eea

\subsection{vacuum diagrams in Schwarzschild geometry \la{vds}}

In this subsection we analyze the loop-generated cosmological constant for a Schwarzschild background. We consider the scalar sector for this part of the demonstration. In addition to serving as an example of the quantum-induced cosmological constant, the Schwarzschild background provides a setup for displaying the effects potentially indicative of an unsmooth horizon.

One of the observations below is that the Schwarzschild geometry vacuum diagram produces the same results  as the flat case when it comes to the coefficient of the cosmological constant term. In the flat case the single particle spectrum is parameterized by a momentum vector square, $k^2$. Since the spectrum can be determined in the asymptotic region, the Schwarzschild geometry should have the same spectrum as the flat case (see, e.g., \cite{Sakurai}\cite{Garcia-Recio:2013ixa}). 
The following steps are exactly analogous to the steps taken in the flat spacetime analysis in \cite{Weinberg2}. The only difference is the basis used in each case to evaluate the trace: in the flat case the plane waves are used whereas in the Schwarzschild it is the basis obtained in the following manner. 

Usually the mode expansion is done with the solutions of 
\bea
\nabla^2 f =0
\eea
In the case of the Schwarzschild background, a detailed study of this equation was conducted in \cite{Jensen:1985in}.
It is possible to construct the Green's function, $G$, by following the standard procedure. Now consider
\bea
g=f-k^2 \int G g
\eea
The mode $f$ may be taken as the plane wave for an asymptotic spacetime.
The mode $g$, which will be used to compute the trace associated with the vacuum one-loop diagram, satisfies 
\bea
\nabla^2 g =-k^2g
\eea
For the one-loop vacuum diagram, let us analyze the following equation:
\bea
i\G_{1-loop}=-\fr12 \ln \det \Big(\fr{iK}{\pi}\Big)=-\fr12 \Tr\ln\Big(\fr{iK}{\pi}\Big)
\eea
where $K$ is basically the kinetic operator, $K_{x,y}\equiv \nabla^2 \d^4(x-y)$.
One can use the modes $g$ above and evaluate the trace. The result is the same as the flat case as we will now examine.
Let us keep things slightly more general and consider the modes that satisfy, in a schematic notation,
\bea
\nabla^2 g_n= -\w_n^2\, g_n;
\eea
one gets
\bea
i\G_{1-loop}=-\fr{1}{2} \sum_n \ln \Big[\fr{i}{\pi} (\w_n^2+m^2) \Big]
\eea
For an asymptotically flat background, it should be possible to go to a basis with $\w_n^2=k^2$. Define
\bea
\mathscr{I}\equiv -\fr{i}{2(2\pi)^4}\int d^4k\; \ln \Big[\fr{i}{\pi} (k^2+m^2) \Big]
\eea
To evaluate this, consider
\bea
\mathscr{I}'''=-\fr{i}{(2\pi)^4} \int d^4k \;\fr1{(k^2+m^2)^{-3}} 
\eea
In dimensional regularization, this leads to
\bea
\mathscr{I}=A+\fr{\G(1-\w)}{(4\pi)^\w}  m^4  \la{scri}
\eea
where $A$ is an undetermined constant.
As previously mentioned and verified through the heuristic steps above, the coefficient of the cosmological constant term (i.e., the coefficient of $\sqrt{g}$) is independent of the black hole mass for the Schwarzschild case: the result is the same as the flat case computation. (In general, however, the coefficients will depend on the parameters of the curved spacetime.)

A few words on the renormalization are in order. It should be natural to the renormalization scheme in which the coefficient of the $\sqrt{-g}$ including the contribution from the minimum of the matter potential is the physical value of the cosmological constant.
With such a scheme the renormalized cosmological constant and the finite parts of the vacuum diagrams should be kept small enough to not overshoot the small value\footnote{See the discussion in \cite{Park:2016zgt} and the refs therein on the potential origin of the smallness of the cosmological constant} of the observed cosmological constant which must be the combined effect of the classical value (if it is non-vanishing) and quantum corrections.

\vspace{.2in}

For the remainder of this subsection, we will take up several issues on the quantum-induced deformation of the horizon-vicinity before getting to the time-dependent background in section 3. The discussion here will naturally motivate the use of a time-dependent solution. First, we explain the reason we have chosen the scalar kinetic term of the stress-energy tensor, as opposed to the Riemann tensor-containing terms generated from the one-loop, for the purpose of examining the energy. Secondly, we outline the steps that lead one to see the deformation of the geometry of the horizon-vicinity.

It has been shown in \cite{Park:2016vam} that the 1PI action at the first few leading orders is given by\footnote{Although the results were obtained for an AdS background, they should be generically valid (see also \cite{Park:2016zgt} where other backgrounds were considered): different backgrounds will, in general, give different values of the coefficient, $e$'s.}
\bea
&&S=\fr1{\k^2}\int d^4 x \sqrt{-g} (R-2\L) 
-\int d^4 x \sqrt{-g}\Big[\fr12(\pa_\mu \f)^2 +{\cal V}(\f)\Big] \nn\\ 
&&\hspace{-.3in}+\fr1{\k^2}\int d^4 x \sqrt{-g}\Big[e_1{ \k^4} R\f^2+e_2 \k^2R^2+e_3\k^2 R_{\m\n}R^{\m\n} +\cdots
\Big]  
\la{qsacttwo} 
\eea
where the subscript `$r$' (denoting `renormalized' quantities) has been omitted.  
The coefficient $e$'s are to be determined by divergence-cancelling conditions. The quantum-corrected stress-energy tensor that follows from this action is
\bea
T_{\m\n}&=&  \pa_\m \f \pa_\n\f-2e_1{ \k^2}R_{\m\n}\f^2+2e_1 { \k^2}\nabla_\m\nabla_\n \f^2     \nn\\
&&+g_{\m\n}\Big[\Lambda -\fr12 (\pa_\mu \f)^2 - {\cal V}(\f)  
+e_1 { \k^2}R\f^2 -2 e_1{ \k^2}\nabla^2\f^2  \Big] +\cdots \nn\\
&=&  \pa_\m \f \pa_\n\f   +g_{\m\n}\Big(\Lambda -\fr12 (\pa_\mu \f)^2 - {\cal V}(\f)  \Big) \\
&&+\k^2e_1 (-2 R_{\m\n}\f^2+2 \nabla_\m\nabla_\n \f^2) +e_1\k^2 g_{\m\n}(
R\z^2 -2 \nabla^2\f^2  ) +\cdots  \nn
\eea
The higher-derivative Riemann tensor-containing terms (such as, e.g., $RR_{\m\n},R_{\m\k_1\k_2\k_3}R_\n{}^{\k_1\k_2\k_3}$ etc) have been excluded from the definition of the stress-energy tensor above. Although they may potentially lead, once contracted with the four-velocity vectors, to divergence in the energy, we have explicitly checked that at one-loop, those terms do not in any case yield divergence for a Schwarzschild or dS Schwarzschild background. The other candidates are the matter sector terms, and our focus in the next section will be the scalar kinetic term. The advantage of considering a {\em time-dependent} black hole solution is that it allows one to consider the kinetic term of the scalar field instead of conducting two- and higher- computations to see the potential divergence in the energy from the graviton sector.

Finally, let us examine the quantum effects on the bulk geometry, in particular, in the horizon vicinity. A thorough understanding of the deformation of the horizon vicinity would require a separate investigation of its own. For example, the definition of the horizon ``vicinity" should be more precisely quantified.\footnote{Based on the analysis in \cite{Park:2013rm} (Eq. (55) therein), we expect it is the region within a coordinate distance of order $R_S$, where $R_S$ denotes the Schwarzschild radius, from the event horizon. This seems in line with other estimations in the recent literature \cite{Almheiri:2012rt,Giddings:2017mym,Dey:2017yez}. See also the analysis towards the end of section 3.} Here we only outline the analysis needed.

It was shown in \cite{Park:2016vam} that the renormalization procedure requires a metric field redefinition of the form
\bea
g_{\m\n}&\ra&  g_{\m\n}+\k^2\Big[l_0g_{\m\n}+l_1  g_{\m\n}R+l_2 R_{\m\n}+l_3 g_{\m\n}(\pa\f)^2+l_4 \pa_\m\f \pa_\n\f+l_5g_{\m\n}\f^2\nn\\
&&+\k^2\Big(l_6R\pa_\m\f \pa_\n \f+l_7R_{\m\n}(\pa\f)^2+l_8R_{\m\n}\f^2+l_9g_{\m\n}R(\pa\f)^2 \nn\\
&& \;\;\;\hspace{.5in}+  l_{10} g_{\m\n}R^{\a\b}  \pa_\a \f\pa_\b \f+l_{11}R^{\a}{}_{\m\n}{}^{\b}  \pa_\a \f\pa_\b \f+l_{12} g_{\m\n}R\f^2\Big) \nn\\
&&\hspace{-.8in}+\k^4\Big(l_{13} g_{\m\n}(\pa \f)^4+l_{14} \pa_\mu \f\pa_\n \f(\pa\f)^2+l_{15} g_{\m\n}\f^4
+l_{16} g_{\m\n}(\pa\f)^2\f^2 +l_{17} \pa_\m\f \pa_\n\f \f^2\Big)+\cdots\Big]\nn\\  \la{msg}
\eea
where the coefficient $l$'s are to be determined so as to absorb the counterterms in \rf{qsacttwo}. For example, the counterterms shifts the graviton sector to
\bea
&&\hspace{1.5in} \fr1{\k^2}\sqrt{-g}R-\fr1{\k^2}2\L \sqrt{-g}\nn\\
&&\hspace{-.3in}\Rightarrow
\sqrt{-g}\Big[-2\Big(\fr{\L}{\k^2}+\fr{\d\L}{\k^2}-2\fr{\d\k \L}{\k^3}+2l_0\k^2 \L\Big)+ \Big(\fr1{\k^2}-2\fr{\d\k }{\k^3}+l_0-\L (4l_1+l_2)\Big)\,R\nn\\
&&\hspace{2in}+\mbox{higher derivative terms} \Big]
\eea
(The details can be found in \cite{Park:2016vam}.) The shifted cosmological constant $\fr{\L}{\k^2}+\fr{\d\L}{\k^2}-2\fr{\d\k \L}{\k^3}+2l_0\k^2 \L$ and the coefficient of $R$ can fixed according to one's renormalization scheme; the renormalization conditions will then determine $\d\L$ and $\d\k$ in terms of the other quantities.  Again, although this result was obtained for a certain fixed background, we do not expect that the generic form of the field redefinition above will be sensitive to the background under consideration. 
Suppose that the divergent terms have been removed by the metric field redefinition above. Then there remain the finite terms, and they are in terms of the original metric. Since they are part of the quantum corrections, they come with the coefficients in a higher power of the Newton's constant. Therefore the original field can be replaced by the new fields at the leading order of the quantum correction. At this point there are two approaches that one can take to study the deformation implied by the field redefinition.\footnote{The issues with the boundary conditions have been discussed in  \cite{Park:2016vam} and will be set aside.} One may try to solve, in a brute-force manner, the field equations in terms of the newly-defined metric field. The analysis in the next section is in the spirit of this approach. The other approach - which is entirely technically motivated - is to introduce the metric field redefinition in such a way as to absorb not only the divergences but also the finite terms, thereby casting the quantum-corrected action into the classical form. (In general it will not be possible to cast the 1PI action into the classical form. However, this should be possible at the low orders of quantum and derivative expansion.) Since the solution of the classical form of the action is already known, the quantum-corrected solution may be iteratively found from the metric field redefinition.
Either way it will be possible to quantitatively examine the meaning of the deformation once the solution of the quantum-corrected field equations is obtained.

\section{Loop-induced non-smooth horizon}

In the previous section we have illustrated the generation of the cosmological constant by taking the flat and Schwarzschild cases.
Although it would be ideal to directly extend the analysis to the background of \cite{Chadburn:2013mta}, such a task would involve highly nontrivial technical analysis. 
For example, the analysis analogous to \cite{Jensen:1985in} would have to be preceded before attempting to construct the Green's function, another demanding task in itself. In order to find the quantum-generated cosmological constant with the background of \cite{Chadburn:2013mta} (or any other general curved background), what one should do is to calculate the vacuum diagram by using the propagator associated with the $(g_{{}_B\m\n}+g_{0\m\n})$-field. That diagram can, in turn, be  calculated perturbatively in a $g_{{}_B\m\n}$ series, and in the leading order of $g_{{}_B\m\n}$ one can focus on the diagram with no $g_{{}_B\m\n}$ field appearing in the external lines. In other words, for the purpose of deducing the cosmological constant, one evaluates just the ``vacuum" diagram with propagator associated with only the $g_{0\m\n}$ field; even for that goal one will have to employ a perturbation series in terms of the curved space parameter(s). In spite of all these complications, it is expected at the end that the quantum effects will generate the cosmological constant term in a similar manner as in the previous section.
Thus let us consider, after setting the higher derivative corrections aside,
\bea
&&S=\fr1{\k^2}\int d^4 x \sqrt{-g} R 
-\int d^4 x \sqrt{-g}\Big[\fr12(\pa_\mu \f)^2 +W(\f)\Big] 
\eea
where
\bea
W\equiv {\cal V}+2\fr{\L}{\k^2}
\eea
with ${\cal V}$ the Higgs type potential in \rf{potqm}:
\bea
{\cal V} &=& \fr{\l}{4}\Big(\f^2+\fr{1}{\l} \m^2\Big)^2   \label{potqmq}
\eea
It is important to note that the cosmological constant term $\L$ has a quantum-field-theoretic origin: the cosmological constant $\L$ denotes the combination of the renormalized value and finite part of the one-loop vacuum diagram (with a certain renormalization scheme). Therefore, $\L\sim \k^2$ and it is of the same order as the matter part of the Lagrangian.

In the leading order of the quantum corrections, the solution of the action above can be substituted into the stress-energy tensor so obtained, which is regular at the horizons as highlighted in \cite{Chadburn:2013mta}, and one can focus on the scalar kinetic term. Finally, with the velocity vectors contracted with the resulting onshell stress-energy tensor, one finds a trans-Planckian energy measured by an infalling observer.
We start by reviewing some of the results of \cite{Chadburn:2013mta} in our convention. (For example, we use mostly plus signature.)

\subsection{time-dependent black hole solution}

Let us briefly review the background obtained in \cite{Chadburn:2013mta} by focusing on the event horizon (instead of the cosmological horizon therein illustrated). We give detailed steps because, compared with \cite{Chadburn:2013mta}, our calculation below yields different numerical factors in several places. The general form of the metric with spherical symmetry is
\bea
ds^2=4c \,e^{2\n} B^{-1/2}du dv+B d\Omega^2
\eea
where $c$, with other constants $\r,\a,\s$ to be introduced shortly,\footnote{Below they are determined to be
\bea
c=-8, \quad \s=\fr12, \quad \r=4,\quad \a=8
\eea
} is a constant to be determined.
The field equations that follow from the action are given by
\bea
\f_{,uv}&=& c\,\,W_{,\f}B^{-1/2}e^{2\n}-\fr1{2B}(B_{,u}\f_{,v}+B_{,v}\f_{,u}) \nn\\
B_{,uv} &=& 2c \;(-K B^{1/2}W+B^{-1/2})e^{2\n}  \nn\\
\n_{,uv}&=& -\fr{c}2 \;(K W B^{-1/2}+B^{-3/2})e^{2\n} -\fr{K}2 \f_{,u}\f_{,v}\nn\\
B_{,vv} &=& 2\n_{,v}B_{,v}-K B\f_{,v}^2 \nn\\
B_{,uu} &=& 2\n_{,u}B_{,u}-K B\f_{,u}^2  \la{feom}
\eea
where 
\bea
K\equiv 8\pi G=M_P^{-2}
\eea
These equations can be solved perturbatively in a ``slow roll" approximation.  
In the leading order, the scalar field takes a constant value, and the last two field equations can be combined to yield
\bea
2\n= \ln B_{,v}+\ln G'(u)= \ln B_{,u}+\ln F'(v)
\eea
where $F$ and $G$ are arbitrary integrations functions to be determined and the prime denotes differentiation with respect to the argument of the field. These equations imply that $B$ is a function of $F+G:$$B=B(F+G)$, and
\bea
e^{2\nu}=F'(v)G'(u)B'(F+G) \la{nu}
\eea
Upon being substituted into the second equation of \rf{feom}, this leads to
\bea
B'=4H^2 B^{\fr32}-4 B^{\fr12} +8GM  \la{Bp}
\eea
where 
\bea
H^2\equiv \fr13\k W_0 \quad,\quad W_0\equiv W(\f_0)
\eea
The result above implies
\bea
B' &=& -4B^{1/2}N(B^{1/2}) \nn\\
N(r) &\equiv & 1-\fr{2GM}{r}-H^2 r^2  \la{BpN}
\eea
Introducing the tortoise coordinate 
\bea
r^* \equiv 2\s\int \fr{dr}{N(r)}
\eea 
and the following definitions
\bea
t-r^*={2\r }G(u)\quad,\quad t+r^*=-{ 2\r}F(v)
\eea
where $\s, \r$ are constants to be determined, one can show
\bea
\int \fr{dr}{N(r)}=-{ \fr{\r}{2\s}} \int \fr{dB}{B'}=-{ \fr{\r}{2\s}}  (F+G)
\eea
The functions $F,G$ can be chosen as follows
\bea
F(v)=-\fr1{{\a} N'(r_h)}\ln[N'(r_h)v]  \quad,\quad  G(u)=\fr1{{ \a}N'(r_h)}\ln[-N'(r_h)u]
\eea
where $\a$ is another constant to be determined and $r_h$ denotes the location of the event horizon.
By inverting the relations between $t,r^*$ and $F,G$, one gets
\bea
v=R_h\exp\Big({ \fr{\a}{2\r}}\fr{t+r^*}{R_h}\Big) \quad,\quad u=-R_h\exp\Big({ \fr{\a}{2\r}}\fr{t-r^*}{R_h}\Big)
\eea
where
\bea
R_h\equiv \fr1{N'(r_h)}  \la{Rh}=\fr{r_h}{1-3H^2r_h^2}
\eea
($R_c, R_N$ that appear below are similarly defined.)
With these, \rf{nu} becomes
\bea
e^{2\nu}= -{\fr{c}{\a^2}} \fr{R_h^2 rN}{uv}
\eea
In the leading order the second field equation in \rf{feom} leads to the following relation to the parameters:
\bea
c^2=\fr{\r^2}{\s^2}  \la{crb}
\eea
Let us now extend the analysis to the next order of the approximation of the slowly varying fields. The equations from \rf{Bp} through \rf{Rh} remain valid. 
The field equations in \rf{feom} can be solved perturbatively by taking the following ansatze:
\bea
\f &=& \f_0+\sqrt{2\e}\;M_P \f_1(u,v) \nn\\
B&=& r^2(1+ \e\d_1(u,v))  \nn\\
\n &=& \n_0(u,v)+\e \d_2(u,v)
\eea
where $\f_0$ denotes the constant. The slow roll parameter $\e$ is defined by
\bea
\e=\fr{M_P^2 W'(\f)^2}{2W^2(\f)}  \la{eW}
\eea
With these substituted, the field equations now take
\bea
\f_{1,uv} &=& \fr{N}{uv}\Big[-3{ \fr{c^2}{\a^2}} H^2 R_c^2+{ \fr{\r}{\a\s}}\fr{R_c}{r}(v\f_{1,v}-u\f_{1,u}) \Big] \nn\\
r^2 \d_{1,uv}
&=&  -{ \fr{\r}{\a\s}}\fr{2R_c rN}{uv}(u\d_{1,u}-v\d_{1,v})\nn\\
&&+\fr{R_c^2 N}{uv}\Big[ -4{ \fr{c^2}{\a^2}}\d_2(1-3H^2r^2)+3{ \fr{\r^2}{\a^2\s^2}}(-H^2r^2+1)\d_1+12{ \fr{c^2}{\a^2}}r^2 H^2 \f_1^2 \Big] \nn\\
r^2\d_{2,uv}&=&{ \fr{c^2}{\a^2}}  \fr{R_c^2 N}{uv}\Big[(3H^2r^2+1)\d_2 -\fr34 (H^2r^2+1)\d_1  + 3H^2r^2\f_1 \Big] 
     -r^2 \f_{1,u}\f_{1,v}\nn\\
r^2\d_{1,vv}&=& 4{ \fr{\r}{\a\s}} R_c \fr{rN}{v}(\d_{2,v}-\d_{1,v}) +\fr{\d_{1,v}}{v}\Big({ \fr{\r}{\a\s}}r R_c \fr{\pa (rN)}{\pa r} -r^2\Big) -2 r^2\f_{1,v}^2\nn\\
r^2\d_{1,uu}&=&-4{ \fr{\r}{\a\s}} R_c \fr{rN}{u}(\d_{2,u}-\d_{1,u}) -\fr{\d_{1,u}}{u}\Big({ \fr{\r}{\a\s}}r \Rt_c \fr{\pa (rN)}{\pa r} +r^2\Big) -2 r^2\f_{1,u}^2
\la{1eom}
\eea
Other than the signature difference, these results are the same as the corresponding equations in \cite{Chadburn:2013mta} because $\fr{c^2}{\a^2}=1, \fr{\r}{\a\s}=1$ as we will shortly see. We focus on the first equation above, the scalar field equation. In the $(t,r^*)$ coordinates, the first equation in \rf{1eom} takes
\bea
\ddot{\f}_1-\fr{\pa^2 \f_1}{\pa r^{*2}}+\fr{8}{c}\fr1{r^2}\fr{\pa r^2}{\pa r^{*}}\fr{\pa \f_1}{\pa r^*}
=\Big(\fr{4c}{\r^2}\Big)\;3NH^2
\eea 
Upon choosing
\bea
c=-8
\eea
the equation above can be put into the standard form:
\bea
\ddot{\f}_1-\fr1{r^2}\fr{\pa }{\pa r^{*}}\Big(r^2\fr{\pa {\f}_1}{\pa r^*}\Big)
=3NH^2
\eea 
Taking solution of the scalar field equation as
\bea
\f_1=L\, t+\vf(r)  \la{f1ans}
\eea
one can show that the scalar field equation reduces to
\bea
\fr{d}{dr}\Big[r^2N(r)\fr{d}{dr}\Big( \tilde{\vf} \Big)\Big]+3H^2r^2=0 \la{vf}
\eea
where we have defined
\bea
\tilde{\vf} \equiv \fr{\r^2}{16\, {c}\,\s^2} \;{\vf}
\eea
One can show that \rf{vf} admits the following form of the solution:
\bea
\tilde{\vf}=-\sum_i \Big(H^2r_i+\fr{C}{r_i^2}\Big)R_i \ln |r-r_i|+\fr{C}{2GM}\ln r  \la{vftsol}
\eea
where $i=c,h,N$. ($r_c, r_N$ denote the cosmological constant and negative root of $N(r)=0$, respectively.)
Now we turn to the regularity of the scalar solution.
 The regularity of $\pa_V \f $ and $\pa_u \f$ at the cosmological horizon and event horizon
yields the following two conditions for $C,\tilde{L}$: 
\bea
{ \tilde{L}}-H^2 r_c-\fr{C}{r_c^2}=0={ \tilde{L}}+H^2 r_h+\fr{C}{r_h^2}
\eea
where
\bea
{ \tilde{L}}= \fr{\r^2}{8c \s} L
\eea
from which the integration constants $\l$ and $C$ can be determined as
\bea
{ \tilde{L}}=\fr{r_c-r_h}{r_h^2+r_c^2} \quad,\quad C=-H^2 \fr{r_h^2 r_c^2 (r_h+r_c)}{r_h^2+r_c^2}  \la{Cl}
\eea
Making use of these relations (and the ones among $r_h,r_c,r_N$ in the appendix of \cite{Chadburn:2013mta}), the scalar field solution can be written
\bea
\f&=&\f_0+ \sqrt{2\e}\;M_p L \Big\{ t - R_c \ln |r-r_c|+  R_h\ln (r-r_h)\nn\\
&&+\fr{1}{2} (\fr{r_c}{r_h} R_h-\fr{r_h}{r_c}R_c)\ln(r-r_N)
-\fr{r_hr_c}{r_c-r_h}\ln r \Big\}  \la{fsol}
\eea
For reasons that will become clear below we choose
\bea
\s=\fr12
\eea
which on account of \rf{crb}, implies
\bea
\r=4
\eea
Explicitly seeing the regularity of the scalar field at the event horizon is a matter of rewriting \rf{fsol} as
\bea
\f&=&\f_0+ \sqrt{2\e}\;M_p L \Big\{R_h\ln\fr{v}{R_h} -2R_c \ln |r-r_c|\nn\\
&&-\fr{r_h}{r_c}R_c \ln(r-r_N)
-\fr{r_hr_c}{r_c-r_h}\ln r \Big\}  \la{fsoltwo}
\eea
for which $\s=\fr12, \fr{\a}{2\r}=1$ have been used.

We now show that the cosmological constant is crucial for the solution's time-dependence (at the order of the perturbation under consideration). Consider \rf{f1ans} with \rf{vftsol}. Suppose $H=0$ as would be implied by the vanishing cosmological constant. The expression for $C$ in \rf{Cl} implies $C=0$. Also, with vanishing $H$, the two locations of $r_h$ and $r_c$ become identical, leading to $\tilde{L}=0$ as can be seen from the expression of $\tilde{L}$ in \rf{Cl}. Therefore, the supposed time- and position- dependent part $\f_1$ comes to identically vanish, and the solution reduces to the usual Schwarzschild black hole.

\subsection{energy measured by an infalling observer}

To the leading order in $\e$, one can use the geodesic in the de Sitter Schwarzschild background since the kinetic term of the scala field already comes with $\e$. At the leading order in $\e$, the metric is given by
\bea
ds^2=-f(r)dt^2+\fr{dr^2}{f(r)}+r^2(d\th^2+\sin^2\th d\f^2)
\eea
where
\bea
f(r)=1-\fr{2M}{r}-H^2 r^2
\eea
For a radial motion one gets
\bea
\Big(\fr{d r}{d\t}\Big)^2&=&{\cal E}_0^2-(1-\fr{2M}{r}-H^2 r^2)\nn\\
\fr{d t}{d\t}&=&\fr{{\cal E}_0}{1-\fr{2M}{r}-H^2 r^2} \la{geo}
\eea
where ${\cal E}_0$ is the energy per unit rest mass (thus mass-dimensionless).
The stress-energy tensor is given by
\bea
T_{\m\n}^{quan}=\pa_{\m}\f\, \pa_{\n}\f-\fr12 g_{\m\n} (\pa \f)^2+\cdots  \la{seh}
\eea
Finally, let us compute the energy, $E$, measured by an infalling observer\footnote{For a fully proper analysis, a systematic renormalization of the cosmological constant with explicit renormalization conditions will be required.}:
\bea
E=T^{quan}_{\m\n}\fr{d x^\m}{d\t}\fr{d x^\n}{d\t}  \la{enerexp}
\eea	
Note that the stress tensor $T^{quan}_{\m\n}$ itself is finite so the slow roll scheme remains valid: it is only after taking the contraction with the four-velocities that one gets trans-Planckian energy.
Let us focus on the first term\footnote{For example, the term $-\fr12 g_{\m\n}(\pa \z)^2 $ yields a finite result once one extends the unit norm condition for the 4-velocity vector to the quantum level. The part in $(\cdots)$ represents either the graviton sector (depending on one's definition of the stress-energy tensor) or higher-order quantum corrections. Although we focus on only one of the kinetic terms, the higher order couplings in \rf{seh} will obviously also contribute to the deformation of the geometry.} in \rf{seh}:
\bea
\pa_{\m}\f\, \pa_{\n}\f \fr{d x^\m}{d\t}\fr{d x^\n}{d\t} &=& \Big(\pa_{t}\f \fr{d t}{d\t} +  \pa_{r}\f \fr{d r}{d\t}\Big)^2 \la{en}
\eea
Substituting 
\bea
\pa_{t}\f &=& \sqrt{2\e}\;M_P L \nn\\
\pa_{r}\f &=&\sqrt{2\e}\;M_p L\Big[ \fr{R_h}{r-r_h}-\fr{R_c}{r-r_c} +\fr12(\fr{r_c}{r_h}R_h-\fr{r_h}{r_c}R_c) \fr1{r}
-\fr{r_cr_h}{r_c-r_h}\fr1{r} \Big]  \nn\\
\fr{d r}{d\t}&=&-\sqrt{{\cal E}_0^2-(1-\fr{2M}{r}-H^2 r^2)}\quad,\quad \fr{d t}{d\t}=\fr{{\cal E}_0}{1-\fr{2M}{r}-H^2 r^2} 
\eea
into \rf{en}, one gets
\bea
\simeq (2\e M_P^2 L^2)\Big[\fr{{\cal E}_0}{N} -\sqrt{{\cal E}_0^2-N}\Big( \fr{R_h}{r-r_h}-\fr{R_c}{r-r_c} +\fr12(\fr{r_c}{r_h}R_h-\fr{r_h}{r_c}R_c) \fr1{r}
-\fr{r_cr_h}{r_c-r_h}\fr1{r} \Big)  \Big]^2  \la{Ple} \nn\\
\eea
Upon taking the $r\ra r_h$ limit and using $\fr1{N(r)}\simeq \fr{R_h}{r-r_h}$ in this limit, the equation above becomes
\bea
\simeq (2\e M^2 L^2)\Big[-{\cal E}_0\Big( -\fr{R_c}{r_h-r_c} +\fr12(\fr{r_c}{r_h}R_h-\fr{r_h}{r_c}R_c) \fr1{r_h}
-\fr{r_cr_h}{r_c-r_h}\fr1{r_h} \Big)  \Big]^2\
\eea
Since $r_c >> r_h, H^2<<1$ for the realistic values of the parameters, the following approximations can be adopted:
\bea
L \sim \fr1{r_c}\quad,\quad R_h \sim r_h\quad,\quad R_c\sim { -\fr12}r_c
\eea
and one gets
\bea
\pa_{\m}\f\, \pa_{\n}\f \fr{d x^\m}{d\t}\fr{d x^\n}{d\t} &\simeq& \fr12 \e M_P^2 \fr{{\cal E}_0^2 }{r_h^2}  \la{enerexp2}
\eea
This shows that the energy measured by an infalling observer is trans-Planckian in line with Firewall. Note that small parameters do not always lead to small effects: the classical cosmological constant piece ${\cal V}(\f_0)$ and quantum piece $\L$ should be small to be realistic. They together (i.e., through $H^2$) determine the locations of the horizons, $r_h, r_c$. The overall smallness of $H^2$ is what makes $r_c$ large whose absence in turn is important in yielding the trans-Planckian energy above. (In contrast, a factor of $\fr1{r_c^2}$, instead of $\fr1{r_h^2}$, will appear in the energy formula for the case of the cosmological event horizon to which we turn below.)

One potential objection is the validity of the perturbation series in $\e$ when its coefficient becomes large. Note that the series is valid if $r$ is large. In this context it will be illuminating to examine two regions: one region with $r\sim (1+s) \,r_h$ where $s$ is a number of order 1 and the other region with $r\ra r_c$. One can show that in the leading order of $H$, the maximum of \rf{Ple} occurs at
\bea
s=\fr{2{\cal E}_0^2(1+2{\cal E}_0^2)}{-1+8{\cal E}_0^2+8{\cal E}_0^4}
\eea
For ${\cal E}_0^2< \frac{1}{4} \left(\sqrt{6}-2\right)\simeq 0.11$, this is a negative number; the energy measured by an infalling observer decreases monotonically in the interval of $r_h<r<r_c$.\footnote{It is curious that for ${\cal E}_0^2>\frac{1}{4} \left(\sqrt{6}-2\right)$, the maximum falls on the location outside of $r=r_h$.} It is also useful to find the location of $r$ where the value of energy becomes $\fr1{e}\simeq 0.37$ of the value at $r=r_h$: again at the leading order of $H$, it is given by
\bea
r= \Big[1+0.63\; {\cal E}_0^2+{\cal O}({\cal E}_0^3)\Big] \;r_h
\eea 
Therefore the energy \rf{enerexp2} decreases rather rapidly with a reasonably small value of $\e$, which seems consistent with the expectation that the quantum deformation is within the order of $R_S$ from the event horizon. For the cosmological horizon the sign of $\fr{d r}{d\t}$ will be opposite:
\bea
\fr{d r}{d\t}&=&\sqrt{{\cal E}_0^2-(1-\fr{2M}{r}-H^2 r^2)} 
\eea
The $\fr1{r-r_c}$ terms in \rf{Ple} cancel and the rest of the terms are small so, unlike the event horizon, the Firewall does not exist in the case of the cosmological horizon.

\section{Conclusion}

There has long been an anticipation that the back reactions must be important in the black hole physics. A solution with a back reaction should naturally be time-dependent. In an interesting work, \cite{Chadburn:2013mta}, a time-dependent solution has been obtained on general grounds, and in this work we have analyzed its quantum deformation of the horizon-vicinity geometry by computing the onshell (and offshell) stress-energy tensor. We have pointed out that the metric field redefinition required in the renormalization procedure \cite{Park:2016zgt} implies the quantum-induced deformation of the geometry.  
We have computed the energy measured by an infalling observer in a time-dependent black hole background. The computation has led to a trans-Planckian energy in line with the Firewall and related proposals.  
The role of the loop-generated cosmological constant was important in consonance with the anticipation in \cite{Park:2013rm} that the quantum corrections will be important in the physics of the horizon vicinity.

The following several facts are worth highlighting and they naturally suggest future directions. Firstly, the role of the cosmological constant in this work is intriguing: given the fact that the measured value is small the quantum contribution should be important regardless of whether there exists an intrinsic classical piece, ${\cal V}(\f_0)$. A more complete understanding of the physics of the cosmological constant including its generation mechanism and systematic renormalization may hold the key to several important problems: in addition to having shed light on the cosmological constant problem itself as well as the black hole information paradox as sketched in the present and recent related works \cite{Park:2016fxc,Park:2016vam}, it may well help accomplish the quantization of gravity even in an {\em arbitrary} background since the presence of the cosmological constant provides a quite flexible leverage. 

Secondly, the present and related works \cite{Park:2016fxc,Park:2016vam} suggest that the information may be stored split over the horizon vicinity and asymptotic boundary region.  
The field redefinition of the metric recently discussed in \cite{Park:2016zgt} signifies a change in the spacetime geometry as has been analyzed in the main body. It will be useful to more thoroughly examine, including the contributions from various other terms in \rf{seh}, how significant is the deformation of the metric over the ranges of order $R_S$ from the horizon. Numerical works by using realistic values of the parameters will be useful along this line of study.

There is a question about whether or not a combined system of, say, a Schwarzschild black hole and an infalling observer or detector can be viewed as truly time-independent. An infalling detector will observe jets emitted from the detector as it disintegrates - a process viewed as ``bleaching" of information in \cite{Park:2013rm,Park:2013bma}. To properly (i.e., quantum-field-theoretically) test such effects, one will have to study how an incoming localized wave-packet behaves as it approaches the horizon. It is expected that even such a small disturbance gets amplified near the horizon and thereby significantly deforms the geometry, making it time-dependent as a matter of principle. If the mass of the infalling observer is small compared with the BH then the overall effect for the black hole will be small at the end. However, the effect will not necessarily be small for the observer himself while the bleaching process is in progress.

Even a slight perturbation in the vicinity of the event horizon will be greatly amplified as the perturbation travels down to the throat and its potential energy gets converted into jet-like emissions. It would be of great interest (see, e.g., \cite{Giddings:2016btb} for a related discussion) to see if such effects could be detected in a near-future endeavor such as the Event Horizon Telescope \cite{Doeleman:2009te}. 
In particular the quantum effects discussed in the present work may perhaps be responsible for the extreme high-energy gamma rays that seem to originate from Sgr A$^*$ at the center of our galaxy.

\newpage
\appendix




\newpage

\end{document}